# Superconductivity in a new type of copper oxide

W.M. Li[1], J.F. Zhao[1], L.P. Cao[1], Z. Hu[2], Q.Z. Huang[3], X.C. Wang[1], Y. Liu[1], G.Q. Zhao[1], J. Zhang[1], Q.Q. Liu[1], R.Z. Yu [1], Y.W. Long[1,8], H. Wu[3], H.J. Lin[4], C.T. Chen[4], Z. Li[5], Z.Z.Gong[6], Z.Guguchia[6],J.S. Kim[7], G. R.Stewart[7],Y.J. Uemura[6]. S.Uchida[1,9], C.Q Jin[1,8]

[1] Institute of Physics; School of Physics, University of Chinese Academy of Sciences, Chinese Academy of Sciences, Beijing 100190, Materials Research Lab at Songshan Lake, Dongguan, China;

2. Max Planck Institute for Chemical Physics of Solids, Dresden, Germany;

3. NIST Center for Neutron Research, Gaithersburg, MD 20899, USA;

4. National Synchrotron Radiation Research Center (NSRRC), 101 Hsin Ann Road, Hsinchu 30076, Taiwan;

5. College of Materials Science and Engineering, Nanjing University of Science and Technology, Nanjing 210094, China;

6. Department of Physics, Columbia University, New York, USA;

7. Department of Physics, University of Florida, Gainesville, FL, 32611-8440, USA

8. Collaborative Innovation Center of Quantum Matter, Beijing100190, China

9. Department of Physics, University of Tokyo, Tokyo 113-0033, Japan

The mechanism of superconductivity in cuprates remains one of the big challenges of condensed matter physics. High $T_c$ cuprates crystallize into layered perovskite structure featuring copper oxygen octahedral coordination. Due to the Jahn Teller effect in combination with the strong static Coulomb interaction, the octahedra in high $T_c$ cuprates are elongated along the $c$ axis, leading to a $3dx^2$-$y^2$ orbital at the top of the band structure wherein the doped holes reside. This scenario gives rise to two dimensional characteristics in high $T_c$ cuprates that favor $d$ wave pairing symmetry. Here we report superconductivity in a cuprate $Ba_2CuO_{4-y}$ wherein the local octahedron is in a very exceptional compressed version. The $Ba_2CuO_{4-y}$ compound was synthesized at high pressure at high temperatures, and shows bulk superconductivity with critical temperature ($T_c$) above 70 K at ambient conditions. This superconducting transition temperature is more than 30 K higher than the $T_c$ for the isostructural counterparts based on classical $La_2CuO_4$. X-ray absorption measurements indicate the heavily doped nature of the $Ba_2CuO_{4-y}$ superconductor. In compressed octahedron the $3d3z^2$-$r^2$ orbital will be lifted above the $3dx^2$-$y^2$ orbital, leading to significant three dimensional nature in addition to the conventional $3dx^2$-$y^2$ orbital.  This work sheds important new light on advancing our comprehensive understanding of the superconducting mechanism of high $T_c$ in cuprate materials.

# Scientific Significance Statement

**Superconductivity is one of the most mysterious phenomena in nature in that the materials can conduct electrical current without any resistance. The cuprates hold the record high superconducting temperature at room pressure so far, but understanding their superconducting mechanism remains one of big challenges. Here we report high Tc superconductivity in $Ba_2CuO_{4-\delta}$ with two unique features: an exceptionally compressed local octahedron and heavily over-doped hole carriers. These two features are in sharp contrast to the favorable criteria for all previously known cuprate superconductors. Thus, the discovery of high Tc superconductivity in $Ba_2CuO_{4-\delta}$ calls into question the widely accepted scenario of superconductivity in the cuprates. This discovery provides a totally new direction to search for further high Tc superconductors.**

\body

A large number of cuprates have been found to show high-$T_c$ superconductivity (HTS) [1–3]. The first high $T_c$ cuprate was $La_{2-x}Ba_xCuO_4$ with a $K_2NiF_4$ (214) type layered structure with a $CuO_2$ plane. Such a $CuO_2$ plane turns out to be a common structural ingredient for all heretofore known high $T_c$ cuprates. The $CuO_2$ plane is stacked with the charge reservoir layer where the carriers are generated to sustain super current in the $CuO_2$ plane. The $CuO_2$ plane is bonded to the apical oxygen atoms at the charge reservoir layer to form octahedral coordination. The distance of an apical oxygen from the in-plane Cu is appreciably longer than the in-plane Cu-O bond length. This is due to the so called Jahn Teller effect plus the interlayer Coulomb interactions which make otherwise degenerate Cu $e_g$ orbitals ($3dx^2$- $y^2$ and $3d3z^2$- $r^2$ orbitals) split, and the topmost $dx^2$- $y^2$ level is well separated from the $d3z^2$- $r^2$ (abbreviated by $dz^2$ hereafter) level. This, together with the strong electronic correlation on the Cu atom, leads to a unique electronic structure of the cuprates. Carriers that are doped into the $CuO_2$ planes by chemical substitution and/or by addition of excess oxygen atoms primarily go onto the $2p_x$ and $2p_y$ orbitals of the in plane oxygen atoms, forming the so called Zhang Rice singlet[4] via strong hybridization with the neighboring Cu$3dx^2$-$y^2$ orbital. This makes the high $T_c$ cuprates effectively a single band system from which superconductivity is thought to emerge with $d$ wave pairing symmetry. These properties form a basis for the elucidation of the pairing mechanism as well as for the exploration of higher $T_c$ [5–7].

Currently consensuses are first, $T_c$ is sensitive to the doping level ($p$) Secondly, a $T_c$-dome forms in the low doping level $p$ region (typically centered around $p \sim 0.15$) which is in proximity to the Mott or antiferromagnetic (AF) insulating phase below $p \sim 0.05$[7] Thirdly, overdoping beyond the dome diminishes superconductivity and the material becomes a Fermi-liquid-like metal in which electronic correlations become weak[7,8]. A heretofore firm correlation between the maximum $T_c$ value in each class and the apical oxygen distance ($d_A$) from the in-plane Cu has frequently been discussed[9,10]. With decreasing $d_A$, the increased contribution of the $dz^2$ orbital to the low lying states near the Fermi level has been argued to weaken pairing interactions and thereby to reduce $T_c$[11,12].

We report here superconductivity in the $Ba_2CuO_{4-y}$ (or equivalently $Ba_2CuO_{3+\delta}$) compound synthesized at extremely high pressures at high temperature. Since the radius of the $Ba^{2+}$ ion is too large to be incorporated in the 214 structure under ordinary conditions, synthesis of bulk materials with a metastable structure by using high pressure was necessary. High pressure synthesized samples of $Ba_2CuO_{4-y}$ show superconductivity with $T_c$ around 73 K, about 30 K higher than that of the isostructural $La_{2-x}Sr_xCuO_4$, the prototypical high $T_c$ cuprate. The present study reveals that this new cuprate has quite unexpected features: (I) the apical oxygen distance can be extraordinarily shorter than that known for all other cuprate superconductors so far; (II) A unique compressed version of the local octahedron becomes available; (III) HTS is realized at very high hole doping level, contrary to the value of p~0.15 discussed above for the previously known high $T_c$ cuprates. All three characteristics have been thought to be unfavorable for high $T_c$ in the previously discovered cuprates[8~19]. Therefore, the present material is a distinctly new kind of high $T_c$ cuprate, and challenges the established wisdom of high $T_c$ superconductivity.

Polycrystalline $Ba_2CuO_{4-y}$ (Ba214) samples are synthesized at high pressure (~ 18 GPa),

much higher than usually used (~ 6 GPa) for the high-pressure synthesis of oxide materials[15,18,19], and at high temperature (~1000˚C) under a highly oxidizing atmosphere. High $T_c$ superconducting samples were produced in the narrow range of the nominal oxygen deficiency $y$ ~ 0.8(excess oxygen content δ ~ 0.2). Shown in **Fig. 1(a)** is the magnetization $M/H$ of a $Ba_2CuO_{4-y}$ polycrystalline sample measured in both zero field cooled (ZFC, shielding) and field cooled (FC, Meissner) modes in a magnetic field of 30 Oe. The sample exhibits a clear superconducting transition at the onset temperature 73 K. The large superconducting volume fraction estimated from *dc* magnetic susceptibility measurements as high as 30% indicated the bulk superconductivity behavior. The conclusion is further supported by muon spin rotation (μSR) and the specific heat measurements. All three measurements guarantee the bulk superconducting phenomenon of the samples. This is fairly large for samples synthesized under high pressure. A high pressure synthesized Ba214 polycrystalline sample is generally composed of very fine grains with submicron size. This results in significant flux penetration at the grain surface which dramatically reduces the Meissner signal[20]. Therefore, the Meissner volume fraction should be regarded as a lower bound of the superconducting volume fraction. This evidence for bulk superconductivity, also confirmed by the muon spin rotation (μSR) showing approximately 40% superfluid density and the specific heat measurements as shown in Fig.1(b), Fig.1(c), respectively, guarantees that the structure measured corresponds to the superconducting phase.

X-ray diffraction was measured for different batches of Ba214 samples to examine the phase purity (a representative XRD pattern is shown in **Fig. 2)**, and is consistent with the $La_2CuO_4$ type structure with space group *I*4/*mmm*. The intensities and shapes of diffraction peaks agree with the previously well characterized high $T_c$ cuprates and the statistics of the pattern is good enough for a detailed structural refinement. Rietveld refinement yields the lattice parameters of the compound with $a$ = 4.003 Å and $c$ = 12.94 Å at room temperature, respectively. The summary of the structure based on Rietveld refinements from powder X ray diffraction patterns is shown in SI Appendix, Table S1. It yields the apical oxygen distance $d_A$ = 1.86 Å. The Cu-O bond lengths for Ba214 at room temperature are estimated to be 2.00 Å in the plane and 1.86 Å along the $c$ axis (corresponding to the apical oxygen distance $d_A$). These values should be taken as average values of the bond lengths. The 2.00 Å in plane Cu-O bond length of Ba214 is the record for the longest among hole doped cuprates, normally ranging from 1.88 to 1.96 Å (see **Fig. 3**)[21-23]. By contrast, the apical oxygen distance $d_A$ = 1.86 Å is the shortest known among the cuprates: about 25 % shorter than the typical value of 2.42 Å in $La_2CuO_4$. The large ionic radius of $Ba^{2+}$ without any other nearby spacer layers in Ba214 expands the in plane Cu-O bond dramatically. Also, it is inferred that the short apical oxygen distance might arise from the electroneutral [$Ba_2O_2$] spacer layer. This neutral [$Ba_2O_2$] layer, without other charge reservoir layers, would allow the apical oxygen to come near the plane, thus realizing the heretofore unprecedented situation that the apical oxygen to Cu bond length is appreciably shorter than the in plane Cu-O bond length.

As in the case of ordinary high $T_c$ cuprates, useful information on the distribution of holes in Cu3$d$ and O2$p$ states can be obtained from the study of soft X-ray absorption spectra at the Cu-$L_3$ edge and the O-$K$ edge[24-26]. In particular, the O-$K$ XAS spectrum provides the number of doped holes quantitatively since the doped holes in cuprates mainly locate at the O2*p* orbitals[24-26]. The O-*K* edge XAS spectrum of Ba214 is presented in **Fig. 4a** together with those of LSCO ($x$ = 0 and 0.15) taken from **Ref. 24**. The weak peak U at higher energy is assigned to the transitions to the upper Hubbard band from the O 1*s* core level (corresponding to the '$Cu^{2+}$' state, for simplicity) which corresponds to the major pre-edge peak in the undoped charge-transfer insulator $La_2CuO_4$. The dominant low energy structure H seen for both Ba214 and LSCO ($x$ = 0.15) is attributable to the transitions from O1*s* to the doped hole states constructed by the strong O2*p* Cu3*d* hybridization (so called Zhang Rice singlet state or '$Cu^{3+}$' state). As demonstrated for LSCO the spectral weight of H(U) increases(decreases) with doping level due to the spectral weight transfer from the peak U to H[24,26]. The spectral weight of U in Ba214 is weaker than that in LSCO ($x$ = 0.15) in Fig.4a indicating a heavily doped phase for our Ba214 sample.

The Cu-$L_3$ XAS spectrum is displayed in **Fig. 4b** together with spectra for the overdoped $La_{2-x}Sr_xCuO_4$ (LSCO, $x$ = 0.34)[25] and perovskite $LaCuO_3$[27] as references. Both of the latter are non-superconducting metals. The spectrum of Ba214 is characterized by two peaks, A and B. The dominant peak A at 931 eV, commonly observed for the three cuprates, is assigned to the transition from a Cu2*p* core level to the electron-empty Cu3*d* upper Hubbard band (from the initial state $2p^6 3d^9$ to the final state $2p^5 3d^{10}$ associated with the nominal $Cu^{2+}$ state). The subdominant higher energy peak B at 932.4 eV is related to the doped holes (nominal $Cu^{3+}$ state) and is attributed to the transition from the $2p^6 3d^9 L$ initial state to the $2p^5 3d^{10} L$ final state ($L$ refers to a hole in the O2*p* ligand state). It is known that the spectral intensity of B at the Cu-$L_3$ edge spectrum is sensitively dependent on the specific arrangement of the Cu-O network[28]. For the corner O shared networks (180° Cu-O-Cu bond) such as those in LSCO and $LaCuO_3$, the intensity of B is strongly reduced due to the strong hybridization between neighboring Cu3$dx^2y^2$ and O2$p_{x,y}$ orbitals which acts to screen the Cu core holes. Because of this effect the feature B is hard to see in the spectrum of LSCO with $x$ = 0.15 and is only seen as a weak high-energy tail for overdoped $x$ = 0.34[25,26]. It appears as a subdominant peak even for 'all-$Cu^{3+}$' $LaCuO_3$ ($p$ = 1)[27]. In the Cu-$L_3$ XAS spectrum of Ba214, the B peak is also subdominant as in the case of $LaCuO_3$ (where the copper is in the extreme high valence of $Cu^{3+}$) , but its intensity is significantly stronger than that for heavily overdoped LSCO (x =0.34) in Ref 25. This further demonstrates a heavily doped phase for the Ba214 sample. The result is not only indicative of the presence of strong Cu-O-Cu bond with bond angle of nearly 180°, but also gives support for a very high doping level in Ba214, consistent with the estimated y values. The combined results of the O-$K$ and Cu-$L_3$ XAS indicate not only that the hole density is fairly high, but also that the doped holes are predominantly on the strongly hybridized Cu-O orbitals, like the Zhang Rice singlet also in the present cuprate.

The longer apical oxygen distance, *i.e.*, an elongated octahedron, generically seen in high $T_c$ cuprates pushes the 3$dx^2$-$y^2$ orbital level above the 3$dz^2$ orbital level. Hence, the doped holes reside primarily on the 3$dx^2$-$y^2$ orbital (or in the Zhang Rice singlet states) that causes the carriers to have predominantly in plane orbital character. To the contrary, a consequence of a shorter $d_A$ is that the 3$dz^2$ orbital level moves above the 3$dx^2$-$y^2$ level as schematically illustrated in **Fig. 3**. This makes the 3$dz^2$ orbital character equally present in the electronic states near the Fermi level with enhanced interlayer coupling and thereby renders this HTS cuprate a multiband system like the iron based superconductors as preliminarily presented in SI Appendix, Fig.S1. Two cuprate superconducting systems have been reported, both characterized as heavily overdoped.

One is $Cu_{0.75}Mo_{0.25}Sr_2YCu_2O_{7.54}$ which is in the heavily overdoped regime ($p \sim 0.46$) with $d_A = 2.165$ Å[20,29] (vs the typical value of 2.42 Å in $La_2CuO_4$ and the 1.86 Å for the Ba214 cuprate reported here). The other is a monolayer $CuO_2$ deposited on a single crystal of $Bi_2Sr_2CaCu_2O_{8+\delta}$[30]. The monolayer $CuO_2$ is thought to be heavily overdoped due to charge transfer at the interface[31]. Density functional theory (DFT) gives a simulated hole density of $p \sim 0.9$ and an apical oxygen distance $d_A = 2.11$ Å with elongated octahedron[32]. Both are supposed to be multiband systems with multiple Fermi surface pockets with $3dx^2\text{-}y^2$ and $3dz^2$ orbital character.

HTS in the present cuprate emerges under apparently unique circumstances: short apical oxygen distance, compressed local octahedron version, and heavily hole overdoped. These properties were thought to be detrimental for high $T_c$ in the previously known cuprates but appear to cooperate to produce HTS in this new type of cuprate. These unusual properties in this new, synthesized at high pressure bulk cuprate superconductor offer important input to theory for understanding of the mechanism of high $T_c$ in cuprate materials in general.

## Figure Captions

**Fig. 1| (a)Magnetization in the superconducting state.** Temperature dependence of magnetic susceptibility (magnetization *M*/*H*) of the $Ba_2CuO_{4-y}$ compound measured in a magnetic field of 30 Oe. Both ZFC and FC mode show a sharp superconducting transition with an onset at $T_c$ = 73 K. **(b)** Superconducting volume fraction in terms of superfluid density estimated from μSR plotted as a function of temperature. **(c)** Temperature dependence of the specific heat measured in the temperature range around $T_c$ on a Ba214 sample for zero and 12 Tesla magnetic applied field.

**Fig. 2| Structural analysis.** Typical X ray ($\lambda$ = 1.54056 Å) powder diffraction pattern of a $Ba_2CuO_{4-y}$ sample measured at room temperature (open circles). The high background in the low angle range is from a covering organic material of Mylar thin film to prevent exposure of the sample to air since the sample is highly hygroscopic. Vertical purple lines indicate the possible Bragg peak positions for the $La_2CuO_4$ type structure with tetragonal symmetry which fit very well to the data as shown by the red solid line. The difference between the observed and calculated patterns is shown by the blue curve at the lower panel of the figure ($R_{wP}$ = 3.41%, $R_P$ = 2.47%, $\chi^2$ = 1.114, where the abbreviations mean respectively weighted profile reliability factor, profile reliability factor, and match factor) evidencing the high quality of the refinement. The lattice parameters thus obtained are $a$ = 4.0030(3) Å and $c$ = 12.942 (1) Å. Numbers in parentheses are standard deviations of the last significant digit.

**Fig. 3| In-plane Cu-O and apical-Cu-O bond lengths.** (upper panel): In-plane Cu-O bond length for various single-layer cuprates: $La_{2-x}Sr_xCuO_4$[21], $Bi_2Sr_2CuO_{6+\delta}$, $Tl_2Ba_2CuO_{6+\delta}$ [22], $HgBa_2CuO_{4+\delta}$ [23], and the present $Ba_2CuO_{4-y}$. (lower panel): The same set of the data for Cu apical-O bond length (apical-O distance). In Ba214 the bond-length ratio is smaller than 1 in which case the $3dz^2$ orbital level is expected to be located above the $3dx^2$-$y^2$ orbital level in contrast to the case where the ratio is significantly larger than 1 as in the case of conventional high-$T_c$ cuprates, sketched in

the right-hand side of the figure. Schematic crystal structure with a compressed 'oxygen octahedron' is also shown (exact positions of oxygen vacancies in the plane are not known at present).

**Fig. 4| XAS and characterization of doped hole states. a:** The O-*K* XAS spectra of Ba214 and $La_{2-x}Sr_xCuO_4$ ($x$ = 0 (blue) and 0.15 (red)) taken from ref. **24**. The background absorption is shown by green lines. The two peaks, U and H, correspond to the transitions from the O1*s* core level to the Cu upper Hubbard band and to the doped hole states, respectively. They are referred to the $Cu^{2+}$ state and the $Cu^{3+}$ state (or Zhang Rice singlet state). **b:** The Cu-$L_3$ XAS spectrum of $Ba_2CuO_{3.2}$ shown together with that for overdoped $La_{2-x}Sr_xCuO_4$ (LSCO, $x$ = 0.34) [25]and $LaCuO_3$[27] as references. The peak A at 931 eV is associated with the transitions from a $2p^63d^9$ initial state to the $2p^53d^{10}$ final state, and the peak B at 932.4 eV is assigned to the transitions from a $2p^63d^9L$ initial state to the $2p^53d^{10}L$ final state ($L$ refers to a hole in the ligand O2*p* state).

**Acknowledgments:**

Works at IOPCAS are supported by MOST & NSF of China through Research Projects(2018YFA0305701; 2017YFA0302901; 11820101003; 2016YFA0300301; 2015CB921000 & 112111KYS820150017). Works at CPS are supported by MPI. Works of Columbia are supported by US NSF DMR1610633, REIMEI project of the Japan Atomic Energy Agency, and a grant from the Friends of University of Tokyo Inc. Foundation. Works at Florida are supported by the United States Department of Energy, Bureau of Energy Sciences (contract. No. DE-FG02-86ER45268). We are grateful to X.H .Chen, J.P. Hu & F.C. Zhang for very helpful discussions. SU wishes to thank the support from CAS for his visit to IOPCAS. CQJ acknowledges T. Xiang, L.Yu, Z.X. Zhao for comments.

**Authors declare there are no conflicts of financial or non financial interests.**

## References:

1. Bednorz JG, Müller KA (1986) Possible high $T_c$ superconductivity in the Ba-La-Cu-O system. *Z Phys B* 64: 189-193.

2. Wu MK, *et al.* (1987) Superconductivity at 93K in a new mixed-phase Y-Ba-Cu-O compound system at ambient pressure. *Phys Rev Lett* 58: 908-910.

3. Zhao ZX, *et al.* (1987) High $T_c$ superconductivity of Sr(Ba)-La-Cu oxides. *Chinese Science Bulletin* 8: 522-525.

4. Zhang FC, Rice TM (1988) Effective Hamiltonian for the superconducting Cu-oxides. *Phys Rev B* 37:3759–3761.

5. Lee PA, Nagaosa N, Wen XG (2006) Doping a Mott insulator: Physics of high-temperature superconductivity. *Rev Mod Phys* 78: 17-85.

6. Scalapino DJ (2012) A common thread: The pairing interactions for unconventional superconductors. *Rev Mod Phys* 84: 1383-1417.

7. Keimer B, Kivelson SA, Norman MR, Uchida S, Zaanen J (2015) From quantum matter to high-temperature superconductivity in copper oxides. *Nature* 518: 179-186.

8. Platé M, *et al.*(2005) Fermi Surface and Quasiparticle Excitations of Overdoped $Tl_2Ba_2CuO_{6+\delta}$. *Phys Rev Lett* 95: 077001.

9. Ohta Y, Tohyama T, Maekawa S (1991) Apex oxygen and critical temperature in copper oxide superconductors: Universal correlation with the stability of local singlets. *Phys Rev B* 43: 2968-2982.

10. Pavarini E, Dasgupta I, Saha-Dasgupta T, Jepsen O, Andersen OK (2001) Band-Structure trend in Hole-Doped cuprates and correlation with $T_{c\ \mathrm{max.}}$ *Phys Rev Lett* 87: 047003.

11. Peng YY *et al*. (2017) Influence of apical oxygen on the extent of in-plane exchange interaction in cuprate superconductors. *Nature Phys* 13: 1201-1206.

12. Sakakibara H, *et al*. (2014) Orbital mixture effect on the Fermi-surface-$T_c$ correlation in the cuprate superconductors: Bilayer vs. single layer. *Phys Rev B* 89: 224505.

13. Yamamoto H, Naito M, Sato H (1997) A New Superconducting Cuprate Prepared by Low-Temperature Thin Film Synthesis in a Ba-Cu-O System. *Jpn J Appl Phys* 36: L341-L344.

14. Karimoto S, Yamamoto H, Sato H, Tsukada A, Naito M (2003) $T_c$ versus lattice constants in MBE-grown $M_2CuO_4$ ($M$=La,Sr,Ba). *J Low Temp Phys* 131: 619.

15. Hiroi Z, Takano M, Azuma M, Takeda Y (1993) A new family of copper oxide superconductors $Sr_{n+1}Cu_nO_{2n+1+\delta}$ stabilized at high pressure. *Nature* 364: 315-316.

16. Chan P, Snyder R (1995) Structure of High Temperature Cuprate Superconductors. *J Am Ceram Soc* 78: 3171-3194.

17. Jarlborg T, Barbiellini B, Markiewicz RS, Bansil A (2012) Different doping from apical and planar oxygen vacancies in $Ba_2CuO_{4-\delta}$ and $La_2CuO_{4-\delta}$. *Phys Rev B* 86:235111.

18. Jin CQ, *et al.* (1995) Superconductivity at 80K in $(Sr,Ca)_3Cu_2O_{4+\delta}Cl_{2-y}$ induced by apical oxygen doping. *Nature* 375: 301-303.

19. Jin CQ, Adachi S, Wu XJ, Yamauchi H (1995) A New Superconducting Homologous Series of Compounds: Cu-12($n$-1)$n$. *Advances in Superconductivity VII*, eds Yamafuji K, Morishita T, (Springer Verlag Tokyo), pp 249-254.

20. Gauzzi A, *et al.* (2016) Bulk superconductivity at 84 K in the strongly overdoped regime of cuprates. *Phys Rev B* 94: 180509(R).

21. Cava RJ, Santoro A, Johnson DW, Jr., Rhodes WW (1987) Crystal structure of the high-temperature superconductor $La_{1.85}Sr_{0.15}CuO_4$ above and below $T_c$. *Phys Rev B* 35: 6716-6720.

22. Torardi CC, *et al.* (1988) Structures of the superconducting oxides $Tl_2Ba_2CuO_6$ and $Bi_2Sr_2CuO_6$. *Phys Rev B* 38: 225-231..

23. Huang Q, Lynn JW, Xiong Q, Chu CW (1995) Oxygen dependence of the crystal structure of $HgBa_2CuO_{4+\delta}$ and its relation to superconductivity. *Phys Rev B* 52: 462-470.

24. Chen CT, *et al.* (1991) Electronic states in $La_{2-x}Sr_xCuO_{4+\delta}$ probed by Soft-X-Ray absorption. *Phys Rev Lett* 66:104-107.

25. Chen CT, *et al*. (1992) Out-of-Plane Orbital Characters of Intrinsic and Doped Holes in $La_{2-x}Sr_xCuO_4$. *Phys Rev Lett* 68: 2543-2546.

26. Pellegrin E, *et al.* (1993) Orbital character of states at the Fermi level in $La_{2-x}Sr_xCuO_4$ and $R_{2-x}Ce_xCuO_4$ (*R*=Nd, Sm). *Phys Rev B* 47: 3354-3367.

27. Mizokawa T, Fujimori A, Namatame H, Takeda Y, Takano M (1998) Electronic structure of tetragonal $LaCuO_3$ studied by photoemission and x-ray absorption spectroscopy. *Phys Rev B* 57: 9550-9556.

28. Hu Z, *et al.* (2002) Doped holes in edge-shared $CuO_2$ chains and the dynamic spectral weight trasfer in X-ray absorption spectroscopy. *Europhys Lett* 59: 135-141.

29. Ono A (1993) High-pressure synthesis of Mo-containing 1212 and 1222 compounds, (Cu, Mo)$Sr_2YCu_2O_z$ and (Cu,Mo)$Sr_2(Y,Ce)_2Cu_2O_z$. *Jpn J Appl Phys* 32:4517-4520.

30. Zhong Y, *et al.* (2016) Nodeless pairing in superconducting copper-oxide monolayer films on $Bi_2Sr_2CaCu_2O_{8+\delta}$. *Science Bulletin* 61: 1239-1247.

31. Zhu G.Y, Wang ZQ, Zhang G.M (2017) Two-dimensional topological superconducting phases emerged from d-wave superconductors in proximity to antiferromagnets. *EPL* 118: 37004.

32. Jiang K, Wu X, Hu J, Wang Z (2018) Nodeless high-$T_c$ superconductivity in heavily-overdoped monolayer $CuO_2$. *Phys Rev Lett* 121: 227002.

33. Jin CQ (2017) Using Pressure Effects to Create New Emergent Materials by Design. *MRS Advances* 2: 2587.

34. Stewart GR (1983) Measurement of low-temperature specific heat. *Rev Sci Instrum* 54: 1-11.

35. Uemura YJ, *et al.* (1989) Universal Correlations between $T_c$ and $n_s/m^*$ (Carrier Density over Effective Mass) in High-$T_c$ Cuprate Superconductors. *Phys Rev Lett* 62: 2317.

36. Junod A, *et al.* (1994) Specific heat up to 14 tesla and magnetization of a $Bi_2Sr_2CaCu_2O_8$ single crystal: Thermodynamics of a 2D superconductor. *Physica C* 229: 209-230.

# METHODS

### Synthesis

In the present work, polycrystalline samples of Ba214 were synthesized using solid state reaction at high pressure and high temperature. First the precursors were prepared by the conventional solid state reaction method from high-purity raw materials BaO and CuO in a molar ratio Ba:Cu:O = 2:1:3. The powder mixture in an appropriate ratio was ground thoroughly in an agate mortar before being calcined at 850 ˚C in an $O_2$ flow for 24 hours with one intermediate grinding. Then, the precursors were mixed with $BaO_2$ and CuO with a molar ratio of 9:2:1 in a dry glove box in order to protect hygroscopic reagents. The role of $BaO_2$ is to create an oxygen atmosphere during the high pressure synthesis of Cu12($n$-1)$n$ homologue series cuprate superconductors as previously described[19]. The samples are synthesized using a so called self oxidization method[33], where the oxidizer itself serves as both chemical composition as well as the atomic oxygen source. The advantage of the method is that it can reduce the unwanted impurity phases implemented from alien oxidizers (such as $KClO_4$). The materials are further subjected to high-pressure synthesis at 18 GPa pressure and at 1000 ˚C temperature for 1 hour with a Walker type multianvil high pressure apparatus, and then quenched to room temperature before releasing the pressure. The 18 GPa pressure was necessary to stabilize the 214 tetragonal phase. The Ba214 tetragonal sample showing a superconductivity onset at $T_c$ of 73 K was obtained by annealing at 150 ˚C for 24 hours under 1 atm. $O_2$ gas flow in a tube furnace.

### Physics Properties Characterization

**Superconducting measurements.** The magnetization measurement is perfomed for the in house characterization of the superconducting state using a Quantum Design VSM facility as shown in **Fig.1a**.

**μSR measurements.** Muon spin rotation (μSR) measruements were performed at TRIUMF in zero field (ZF) and transverse field (TF) with TF = 200 G. The ZF relaxation rate showed a modest increase from 0.1 to 0.35 $\mu s^{-1}$ below $T$ ~ 10 K. This confirms absence of strong magnetism background in the observed TF spectra which exhibit the effect of the superfluid density[34]. The time spectra in TF was fit to signals of two components, a portion of which exhibits a fast damping due to the magnetic penetration depth, and the other component showing temperature independent relaxation due to non-superconducting and paramagnetic volume. The fraction of the superconducting volume shown in **Fig. 1b** was estimated from the amplitude of the former component.

The superconducting volume fraction was estimated by μSR plotted as a function of temperature in terms of superfluid density approximately 40 % at the lowest temperature.

**Specific heat measurements.** In order to avoid the air sensitivity of the sample, the sample was transported sealed in inert gas and coated with GE 7031 varnish in an inert atmosphere glove box before being exposed to the air for less than 5 minutes while being transferred into the calorimeter[35]. This sealing away of the sample below a cured GE 7031 varnish layer appeared to be effective in maintaining the intrinsic properties of the sample. The mass of the sample measured was 53.3 mg; the mass of the cured varnish was 0.54 mg. Note that the $\gamma$ value at the lowest temperature is ~ 14 mJ mole$^{-1}$ K$^{-2}$ which is significantly larger than the value, ~ 3 mJ mole$^{-1}$ K$^{-2}$ for optimally doped YBCO at $T$ = 4.2 K, and is comparable to the values reported for overdoped cuprates[36] which is ascribed to contribution of normal electrons not condensed into the superconducting state.
A characteristic jump-like feature is seen at $T_c$ = 73 K in the temperature dependence of specific heat measured on a Ba214 sample which provides additional evidence for bulk superconductivity (**Fig. 1c**). A crude estimate of the jump $\Delta C$ divided by $T_c$ gives $\Delta C/T_c$ ~ 33 mJ mole$^{-1}$K$^{-2}$, but the values are subject to uncertainty due to possible degradation of the sample during shipping or loading. A jump like feature is also identified at 66 K for an applied magnetic field of 12 Tesla, and the upper critical magnetic field $H_{c2}$ is roughly estimated to be 80 Tesla.

**Structural measurements**

The powder X-ray diffraction is performed basedon on a Rigaku diffractionmeter with $\lambda$ = 1.54056 Å at room temperature. The specimen is covered with transparent organic material (Mylar thin film) to prevent the highly hygroscopic sample from exposure to air. The Rietveld refinement on the Powder X-ray diffraction pattern was performed using the GSAS program. The crystallographic and structural parameters are shown in SI Appendix, Table S1.